\documentclass[usegraphicx]{mn2e}
\usepackage{times}
\newlength{\colwidth}
\begin{document}
\title{The radio galaxy 3C~356 and clues to the trigger mechanisms 
for powerful radio sources}
\author[C.~Simpson and S.~Rawlings]{Chris Simpson$^1$\thanks{Email:
chris@subaru.naoj.org} and Steve Rawlings$^2$\\
$^1$Subaru Telescope, National Astronomical Observatory of Japan, 650
N. A`oh\={o}k\={u} Place, Hilo, HI 96720, U.S.A.\\
$^2$Astrophysics, Department of Physics, Denys Wilkinson Building,
Keble Road, Oxford OX1 3RH}
\maketitle

\begin{abstract}
We present deep near-infrared images, taken with Subaru Telescope, of the
region around the $z=1.08$ radio source 3C~356 which show it to be
associated with a poor cluster of galaxies. We discuss evidence that this
cluster comprises two subclusters traced by the two galaxies previously
proposed as identifications for 3C~356, which both seem to harbour AGN, and
which have the disturbed morphologies expected if they underwent an
interpenetrating collision at the time the radio jets were triggered. We
explain the high luminosity and temperature of the diffuse X-ray emission
from this system as the result of shock-heating of intracluster gas by the
merger of two galaxy groups. Taken together with the results on other
well-studied powerful radio sources, we suggest that the key ingredient for
triggering a powerful radio source, at least at epochs corresponding to $z
\sim 1$, is a galaxy--galaxy interaction which can be orchestrated by the
merger of their parent subclusters. This provides an explanation for the
rapid decline in the number density of powerful radio sources since $z \sim
1$. We argue that attempts to use distant radio-selected clusters to trace
the formation and evolution of the general cluster population must address
ways in which X-ray properties can be influenced by the radio source, both
directly, by mechanisms such as inverse-Compton scattering, and indirectly,
by the fact that the radio source may be preferentially triggered at a
specific time during the formation of the cluster.
\end{abstract}

\section{Introduction}

For a number of years, the $z=1.08$ radio galaxy 3C~356 has been
considered a good test case for understanding the peculiar properties
of distant powerful radio galaxies, and for using such objects to find
and study distant clusters. The peculiar properties of these objects
are typified by the so-called `alignment effect', i.e.\ the tendency
for their optical and radio axes to be aligned (McCarthy et al.\ 1987;
Chambers, Miley \& van Breugel 1987). Studies of 3C~356 provided one
of the first examples of an alignment effect seen at near-infrared
wavelengths (Eales \& Rawlings 1990) and the large (4.8\,arcsec)
separation of the two primary aligned components (denoted `a' and `b'
by Eales \& Rawlings) made it ideal for ground-based studies.
However, the picture is complicated by the fact that both galaxies
display radio emission (Fernini et al.\ 1993) and consequently 3C~356
remains one of the very few 3C objects where the site of the powerful
jet-producing central engine has not yet been unambiguously identified
with a specific host galaxy. Eales \& Rawlings (1990) proposed that
the jet-producing engine was housed in component `b' (brighter in both
the radio and optical), and favoured a model in which component `a'
was a starburst triggered by the advancing radio source. Lacy \&
Rawlings (1994) extended this hypothesis by suggesting that components
`a' and `d' (a further aligned component located on the other side of
`b') had undergone two episodes of star formation driven by
interactions between gas clouds and precessing antiparallel
jets. These ideas were strongly challenged by the detection of broad
emission lines in the polarized-light spectrum of 3C~356a (Cimatti et
al.\ 1997), indicating the presence of a {\em bona fide\/} AGN in this
component.

The idea that powerful radio galaxies can be used to locate distant
clusters of galaxies has its basis in both theory and observations.  The
luminosities of these objects imply the existence of extremely massive
black holes at their centres, which in turn implies the host galaxies
themselves are massive (e.g.\ Haenhelt \& Kauffmann 2000; Laor 2001), a
result supported by observations of radio galaxies at all redshifts. In
addition, the red colours of radio galaxies and the small scatter in the
$K$-band Hubble diagram of their hosts (e.g.\ Eales \& Rawlings 1996;
Simpson, Rawlings \& Lacy 1999) imply that their stellar populations are
old and consequently formed at an early epoch and, in models of structure
formation, galaxies form first in the most overdense regions. 3C~356 was
the first distant radio galaxy to be detected in X-rays (Crawford \& Fabian
1993) and when follow-up observations showed this emission to be diffuse,
it was assumed that the radio source is embedded in a rich cluster
(Crawford \& Fabian 1996). Since then, other distant 3C objects have been
detected (e.g.\ Worrall et al.\ 1994; Crawford \& Fabian 1995), including
the $z=1.786$ radio galaxy 3C~294 (Fabian et al.\ 2001), currently the most
distant intracluster medium detected. Statistical studies reveal an excess
of faint galaxies around the $z \sim 1$ 3C radio galaxies, consistent with
a richness of Abell Class 0 (Best 2000), and therefore similar to the
average found for slightly lower-redshift objects (e.g.\ Hill \& Lilly
1991; Wold et al.\ 2000).  Yet despite the moderate number of X-ray
detections and the statistical evidence, only 3C~324 has had its location
in a rich cluster confirmed (Dickinson 1997), permitting a more detailed
study of its environment (e.g.\ Kajisawa et al.\ 2000a,b). Nevertheless, it
has been suggested (Bremer, Fabian \& Crawford 1997) that the cluster
environment is important in triggering and fuelling the radio source beyond
merely providing a site for a massive elliptical galaxy capable of hosting
a powerful radio source.  The interplay between powerful radio sources and
their environments therefore warrants further investigation, and 3C~356 is
once again a suitable test case for such a study.

In this paper, we present (Section~2) new deep, wide-field
near-infrared images of 3C~356 and new near-infrared spectra of both
the principal components. We combine our imaging data with archival
{\it HST\/} WFPC2 images in Section~3 to investigate the clustering
environment of the radio galaxy. In Section~4, we reconsider the
nature of galaxies `a' and `b'. In Section~5 we discuss the X-ray
properties of 3C~356 and propose a model for the triggering of
powerful radio galaxies, within the framework of which we attempt to
explain the properties of other well-studied powerful radio galaxies.
Finally in Section~6 we provide a brief summary of our conclusions.
Throughout this paper, we adopt $H_0 = 70$\,km\,s$^{-1}$\,Mpc$^{-1}$,
$\Omega_{\rm m}=0.3$, and $\Omega_\Lambda = 0.7$. For this cosmology,
1\,arcsec represents a projected distance of 8.1\,kpc at the redshift
of 3C~356.

\section{Observations and reduction}

\subsection{Deep near-infrared imaging}

\begin{figure*}
\includegraphics[width=\textwidth]{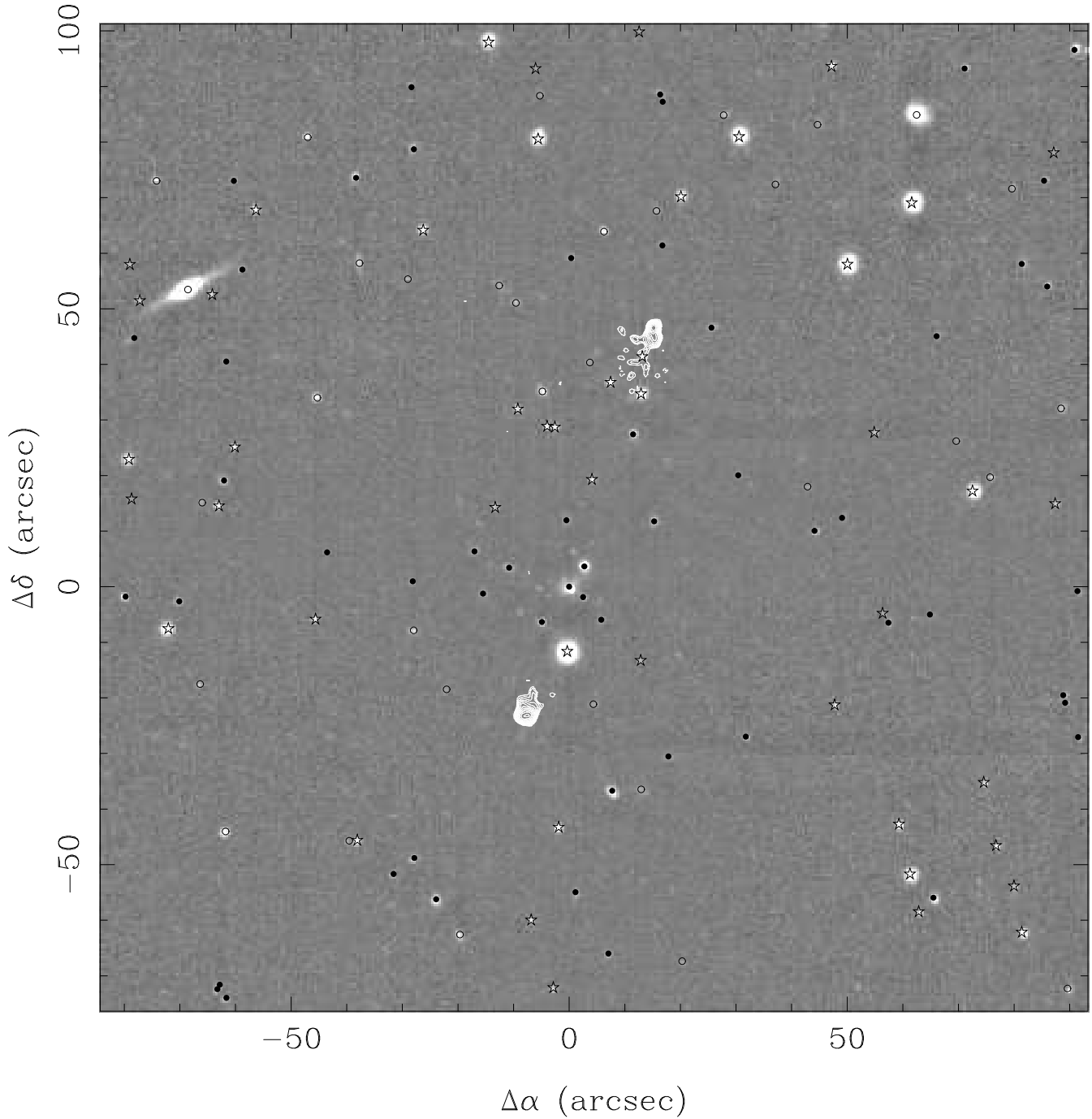}
\caption[]{$K'$-band image of the field of 3C~356. The coordinate
origin is 3C~356b (see text). The greyscale runs between $\pm5\sigma$.
Superimposed (white contours) is the 8-GHz radio map (K.~Blundell, private
communication). Stellar objects are indicated by open stars, while galaxies
are indicated by open or filled circles according to whether they have
$J-K' \le 1.58$ or $J-K' > 1.58$, respectively. Only those objects above
our completeness limit of $K' < 20.8$ are indicated.}
\label{fig:kimage}
\end{figure*}

Deep near-infrared images at $J$ and $K'$ of 3C~356 were obtained with the
Cooled Infrared Spectrograph and Camera for OHS (CISCO; Motohara et al.\
1998), mounted at the Nasmyth focus of Subaru Telescope, on the night of UT
2000 June 17. CISCO has a field of view of $\sim 2$\,arcmin
square. Eight-point jittered mosaics were taken at each of four locations
around the radio galaxy, to cover a square region slightly over 3\,arcmin
on a side.

The data were dark-subtracted and a flatfield created for each
eight-point mosaic from median-filtered, object-masked images. Since
the bias level of CISCO does not stabilize immediately after the
telescope has nodded to a new position, only the last few images at
each position were used to construct the flatfield. An attempt was
made to correct for the variable bias level in the following manner.
The sky level was estimated and an appropriately scaled version of the
flatfield subtracted from the image. An estimate of the residual
signal in each column was then determined and subtracted from that
column, and the final image flatfielded. The success of this method
was determined by fitting a parabola to a log-histogram of the
sigma-clipped pixels to estimate the noise in object-free areas of the
frame, and the procedure repeated iteratively to minimize this noise.
This procedure was extremely successful for the $J$-band data, but
failed to produce an acceptable result for the first $K'$-band image
at each position, and one-eighth of the $K'$-band images had to be
discarded. All the images taken in each filter (which had already been
sky-subtracted by the above procedure) were then mosaicked with bad
pixel rejection, using offsets computed from the centroids of the
brightest unsaturated objects.

The typical exposure times per point were 2560\,s at $J$ and 1120\,s at
$K'$, while the central region received the total exposure time which was
four times greater. The FWHM of the final images were 0.42\,arcsec and
0.55\,arcsec in $J$ and $K'$, respectively. The $J$-band image was aligned
with and then smoothed to the poorer resolution of the $K'$-band image.

The near-infrared images were flux calibrated using the UKIRT faint
standard FS~27 (Hawarden et al.\ 2001). While FS~27 is much bluer than a $z
\sim 1$ galaxy, colour terms for CISCO are negligible compared to the UKIRT
photometric system (F.~Iwamuro, private communication). The systematic
uncertainties in the photometric calibration are $\pm 0.04$\,mag at $J$ and
$\pm 0.03$\,mag at $K'$. With our photometric calibration, the $3\sigma$
limits in a 2-arcsec aperture are $J = 23.56$ and $K' = 21.67$. To ensure a
similar detection threshold across the entire image, we consider only that
part of the images which received at least the exposure times listed
above. The $K'$-band image of this field, measuring 8.436\,arcmin$^2$ and
centered approximately 9\,arcsec north of component `a', is shown in
Fig.~\ref{fig:kimage}.

We also retrieved the {\it Hubble Space Telescope\/} WFPC2 F622W and
F814W images presented in Best, Longair \& R\"{o}ttgering (1997) from
the {\it HST\/} Data Archive (PID 1070; P.I.\ Longair). The images
were combined and sky-subtracted using the {\sc iraf/stsdas} task {\it
crrej\/}. The {\it wmosaic\/} task was used to correct for geometric
distortions so that a simple transformation between the coordinate
frames of the {\it HST\/} and Subaru data could be made. During the
longer F622W exposure, an object moved across the field of view,
leaving a bright track about 2\,arcsec wide. Fortuitously, this
track passed through the photometric aperture of only one source,
which was not affected by cosmic rays in the shorter exposure. We were
therefore able to obtain a photometric measurement for this source,
although the uncertainty is much larger due to the shorter exposure
time.

For consistency with our near-infrared images, we also calibrate the
{\it HST\/} images on the Vega magnitude scale, as opposed to the AB
scale which is more frequently used. The offsets for the F622W and
F814W filters are 0.08 and 0.36\,mag, respectively, in the sense that
$m_{\rm AB} > m_{\rm Vega}$. We do not correct for the small Galactic
extinction ($A_V = 0.06$; Simpson et al.\ 1999).

\subsection{Near-infrared spectroscopy}

We obtained near-infrared spectra of components `a' and `b' on the
night of UT 1999 May 13 using the CGS4 spectrograph on UKIRT. A
0.6-arcsec wide slit was used in conjunction with the
40\,lines\,mm$^{-1}$ grating in second order to provide a resolution
of $\sim 360$\,km\,s$^{-1}$ at the redshifted wavelength of
[O{\sc~iii}]~$\lambda$5007. After peaking up on the bright star to the
south of component `b', the autoguider crosshead was offset 11.3
arcsec south and 0.3 arcsec west and this star was reacquired and used
for guiding during the observation. A position angle of $135^\circ$
ensured that both galaxies `a' and `b' were included within the slit.
A total of 112\,minutes was spent on source, with spectra being taken
at two positions along the slit with the usual `ABBA' technique (e.g.\
Eales \& Rawlings 1993). Bad pixel masking and interleaving of the
$2\times2$ oversampled spectra was performed within {\sc cgs4dr} and
the 2-D spectrum was exported to {\sc iraf} for further reduction. The
flux scale was tied to the UKIRT bright standard HD~162208. A spatial
cut of the standard star spectra revealed a FWHM of 2 pixels (1.2
arcsec), although the true seeing would be less than this due to the
coarse pixel scale. Spectra were extracted along a 3-pixel
(1.8-arcsec) aperture.

\begin{figure}
\includegraphics[angle=-90,width=\colwidth]{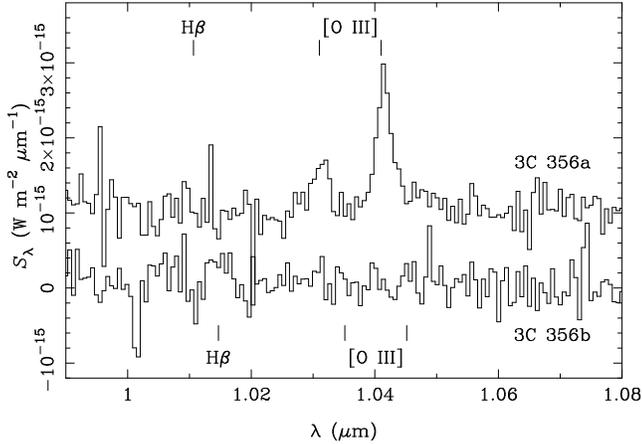}
\caption[]{CGS4 spectra of 3C~356a and b. The spectrum of `a' has been
offset vertically by $10^{-15}$\,W\,m$^{-2}\,\mu\rm m^{-1}$ for
clarity. The locations of the major emission lines are labelled, based
on the observed wavelength of [O{\sc~iii}] in component `a' and the
redshift difference reported in Lacy \& Rawlings (1994).}
\label{fig:spectra}
\end{figure}

The spectra of components `a' and `b' thus extracted are shown in
Fig.~\ref{fig:spectra}. Both components of the
[O{\sc~iii}]~$\lambda\lambda$4959,5007 doublet are clearly detected in
component `a', but no line emission is seen in `b'. The flux of the
$\lambda$5007 line in `a' is $(5.5 \pm 0.3) \times
10^{-18}$\,W\,m$^{-2}$, and it is resolved spectrally with ${\rm FWHM}
\approx 900$\,km\,s$^{-1}$. Assuming a similar width for the narrow
line emission in 3C~356b, we can place a $3\sigma$ upper limit of $7.4
\times 10^{-19}$\,W\,m$^{-2}$ on the strength of this line in our
extraction aperture.

We note that the [O{\sc~iii}]~$\lambda$5007 line flux we measure for
component `a' alone is very similar to that determined by Jackson \&
Rawlings (1997) for `a' and `b' combined. However, Jackson \&
Rawlings' spectrum was taken with 3-arcsec pixels, and therefore they
had extreme difficulty in separating the line emission from the two
components. Our 2-D spectrum clearly shows that all the [O{\sc~iii}]
emission arises in component `a'.

\section{The environment of 3C~356}

\subsection{Source detection and incompleteness}

To investigate the environment of 3C~356, we used SExtractor V2.0.21
(Bertin \& Arnouts 1996). We ran the object detection algorithm on the
$K'$-band image, since this filter samples a rest wavelength of $\sim
1\,\mu$m, and use `two-image mode' to obtain photometry in the other
filters. Photometry was performed in 2-arcsec apertures centered on the $K'$
barycentres. We adopted an object detection threshold of 10 connected
pixels which exceeded the local sky level by $1\sigma$.

We investigated incompleteness in the standard manner of taking isolated
galaxies, reducing their brightnesses, and then adding them back in to the
image at random positions. We estimated the completeness as the fraction of
these sources which are recovered by the detection algorithm. Our catalogue
is 100 per cent complete down to $K' \approx 20.8$ and drops to zero
completeness by $K' \approx 21.6$. The appropriate $k$-correction for a
present-day elliptical galaxy is $-0.73$; we are therefore complete to
$\sim 1.5$\,mag fainter than $M_K^* = -24.63$ (Gardner et al.\ 1997) at $z
\approx 1$. Alternatively, a present-day 13-Gyr-old $L^*$ galaxy would have
$K' \approx 18.1$ at $z=1.08$, accounting for passive evolution according
to the PEGASE models of Fioc \& Rocca-Volmerange (1997), and we should
therefore be able to detect galaxies as faint as $0.1L^*$ at the redshift
of 3C~356.

Since the fractional completeness is not uniform across our image at $K' >
20.8$ due to differences in exposure time, we restrict future discussion in
this paper to those objects which lie above our completeness limit.

\subsection{Star--galaxy separation}

\begin{figure}
\includegraphics[angle=-90,width=\colwidth]{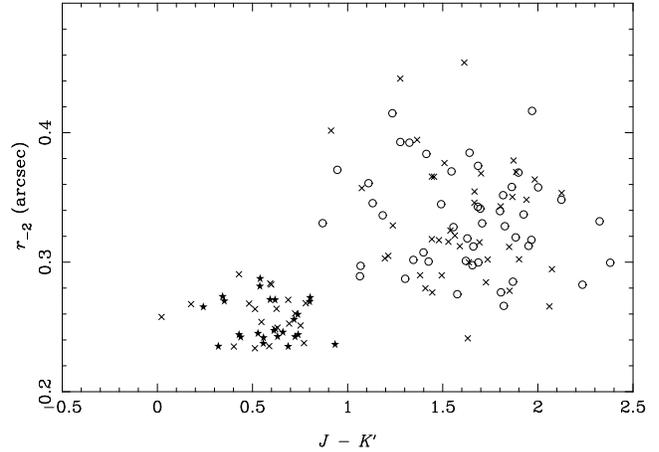}
\caption[]{Plot of $r_{-2}$ (Kron 1980) against $J-K'$ for the 135
unsaturated objects with $K' \leq 20.9$ in the field of 3C~356. Filled
stars and open circles indicate objects which are unresolved and
resolved in the {\it HST\/}/WFPC2 images, respectively, while crosses
indicate objects which do not lie within the WFPC2 field of view.}
\label{fig:stargal}
\end{figure}

Of the 135 objects with $K' \leq 20.8$, 74 lie within the WFPC2 field of
view, and all are readily identifiable in one or both of the {\it HST\/}
images. Of these, we classify 30 as stars based on their having a FWHM of
$\sim 1.5$ pixels or being saturated and displaying diffraction spikes; the
other 44 are all resolved. To classify the remaining 61 objects not within
the WFPC2 field of view, we computed the inverse second moment of the 132
unsaturated objects, $r_{-2}$ (Kron 1980), from the unsmoothed $J$-band
image and plotted this against the $J-K'$ colour in
Fig.~\ref{fig:stargal}. Known stars and galaxies are clearly separated in
this Figure, and we classified the other objects according to their
locations. Our complete sample thus consists of 88 galaxies and 47 stars.

\subsection{The richness of the 3C~356 cluster}
\label{sec:richness}

\begin{figure}
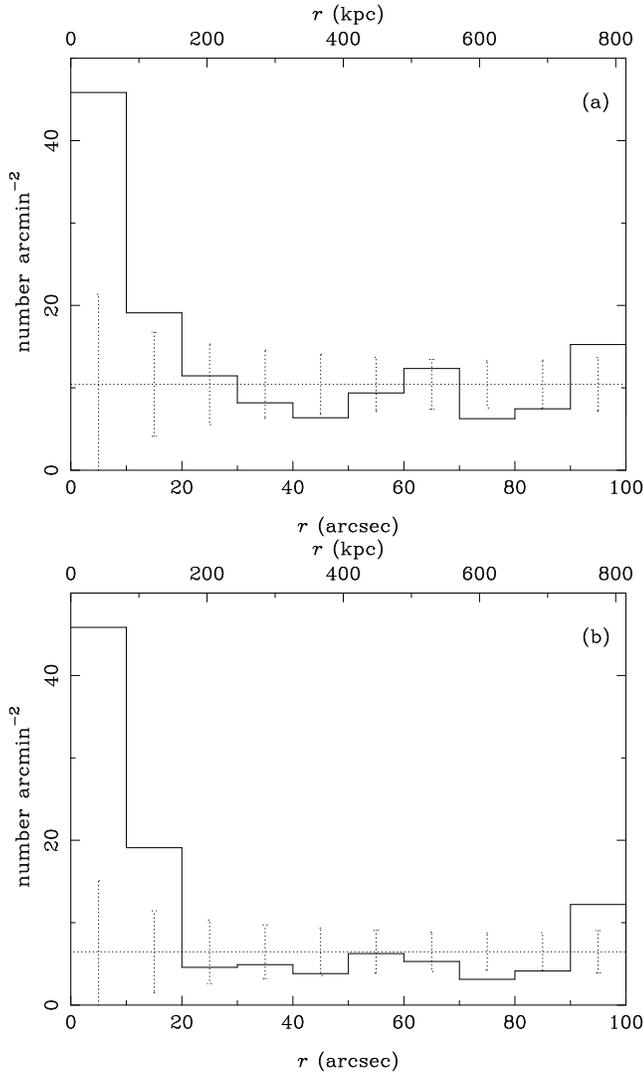

\includegraphics[angle=-90,width=\colwidth]{mb709fig4a.eps}
\includegraphics[angle=-90,width=\colwidth]{mb709fig4b.eps}
\caption[]{(a) Surface density of galaxies as a function of projected
distance from 3C~356b. The dotted line and error bars indicate the
expected distribution if the other 87 galaxies were randomly scattered
about the field. (b) As (a), but for the 54 galaxies with $J-K' >
1.58$.}
\label{fig:radial}
\end{figure}

We note first the small number of galaxies in our field: the number
counts of Gardner, Cowie \& Wainscoat (1993) predict 130 galaxies with
$K < 20.5$ (approximately the same depth as our complete sample after
accounting for aperture losses). This indicates the need to image a
moderately wide field to obtain an accurate field correction when
attempting to detect an excess of galaxies around a high-redshift
object. However, the fraction of red galaxies is large; Elston,
Eisenhardt \& Stanford (in preparation) find that only 44 per cent of
galaxies with $K<20$ have $J-K>1.7$ (S.~A.~Stanford, private
communication), whereas we find the fraction is 56 per cent (assuming
$K'-K=0.1$), inconsistent at the 99 per cent level. In addition, a
plot of the radial distribution of galaxies (Fig.~\ref{fig:radial}a)
clearly shows a highly significant excess of objects around the radio
galaxy. We choose 3C~356b as the origin for investigating a radial
excess as it has the brightest total $K$-band magnitude, and is
therefore presumably the most massive component.

The excess of galaxies near 3C~356 is even more significant when only the
reddest galaxies are considered (Fig.~\ref{fig:radial}b). We classify as
`red' those galaxies with $J-K' > 1.58$; this is the colour of a 4.3-Gyr
old S0 galaxy at $z=1.08$, according to the PEGASE models, and 4.3\,Gyr is
the elapsed time since a plausible formation redshift of $z=5$. While the
probability of 9 randomly selected galaxies all having $J-K' > 1.58$ is
less than 1 per cent, all 9 galaxies within 20\,arcsec of 3C~356b are this
red. We label the objects above our completeness limit in
Fig.~\ref{fig:kimage} with symbols according to their nature.

\begin{figure}
\includegraphics[width=\colwidth]{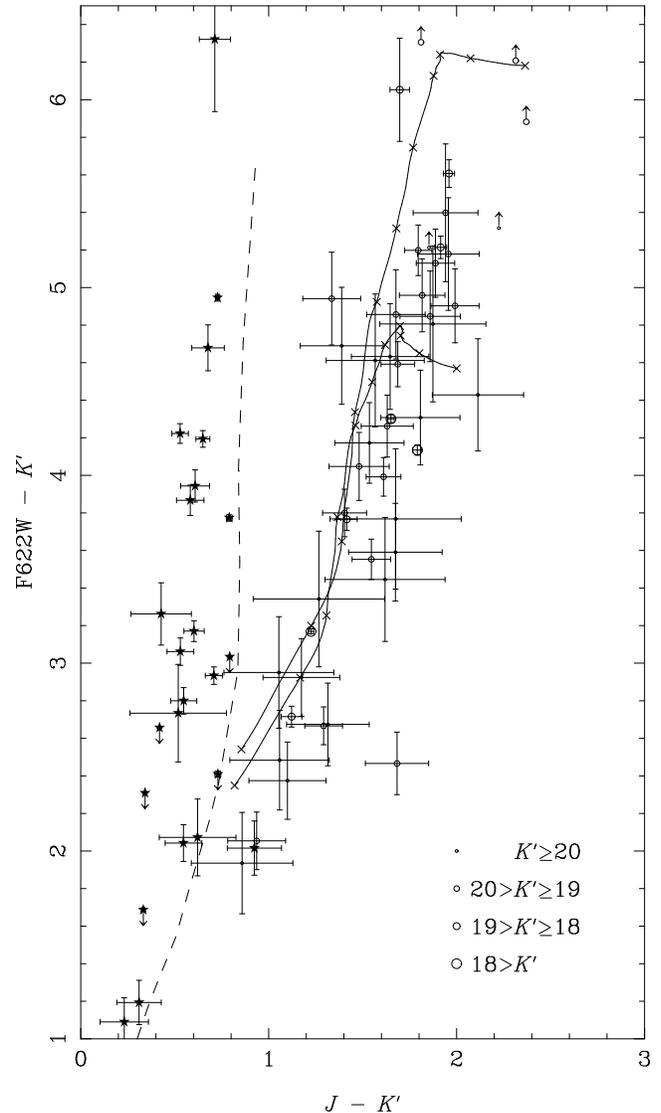}
\caption[]{Colour--colour plot for sources in the field of 3C~356.
Filled stars represent stars, and open circles are galaxies, with the
size of the circle related to the $K'$ magnitude. The dashed line is
the locus of main sequence stars from the spectral library of Fioc \&
Rocca-Volmerange (1997). The solid lines are the loci of 4.3-Gyr old E
and S0 galaxies for $0 \leq z \leq 2$, also from Fioc \&
Rocca-Volmerange. The E locus is the upper of the two, and crosses are
placed at redshift intervals of 0.2.}
\label{fig:colcol}
\end{figure}

We interpret the excess of galaxies around 3C~356, which is visible in
both panels of Fig.~\ref{fig:radial}, as part of a group or cluster
around the radio galaxy. We confirm this in Fig.~\ref{fig:colcol},
where it is apparent that the galaxies which are red in $J-K'$ are
also red in ${\rm F622W}-K'$, and so can only be early-type galaxies
at $z \ga 1$. (It is also clear that our star--galaxy separation has
worked well.)

We estimate the richness of the cluster using Hill \& Lilly's (1991)
$N_{0.5}$ statistic. This is the excess of galaxies within 0.5\,Mpc
(61\,arcsec) of the cluster centre and brighter than $m_1+3$, where
$m_1$ is the magnitude of the brightest cluster member. Hill \& Lilly
discuss the advantages of this method over measurements which rely on
the magnitude of the third brightest cluster member, $m_3$ (e.g.\
Abell 1958; Bahcall 1981). Using our aperture magnitudes, component
`a' is the brighter of the two main components, and hence $K'_1 =
17.87$. Although $K'_1+3$ is therefore slightly fainter than our
completeness limit, we do not apply a correction as there are only 3
galaxies with $20.80 < K' < 20.87$ and we estimate the completeness in
this range to be $>90$ per cent. Although Fig.~\ref{fig:radial}
clearly shows an excess of (red) objects within $\sim 150$\,kpc of the
radio galaxy, this excess is diluted when considering a 0.5-Mpc radius
circle. Based on the mean number density at projected distances
greater than 0.5\,Mpc, we calculate that $N_{0.5} < 14$
($3\sigma$). The shapes of the histograms in Fig.~\ref{fig:radial}
suggest that there are unlikely to be many cluster members at radial
distances greater than 0.5\,Mpc, which would contaminate our
background counts and hence cause us to underestimate the cluster
richness.

We therefore conclude (from table~4 of Hill \& Lilly 1991) that 3C~356
lies in an environment poorer than Abell richness class 0, although
Fig.~\ref{fig:radial} provides compelling evidence that it is not
isolated. This environment is perhaps somewhat poorer than the average
of moderate-redshift radio-loud AGN (e.g.\ Hill \& Lilly 1991; Roche
et al.\ 1999; Best 2000; Wold et al.\ 2000, 2001), but certainly not
unusual given the large spread. However, the environment of 3C~356
represents a much poorer cluster than all 14 of the distant ($z \sim
0.5$) X-ray selected clusters of McNamara et al.\ (2001), which have
similar X-ray luminosities, $L_{\rm X} \sim 10^{37}$\,W, to that
detected by Crawford \& Fabian (1993). There seems therefore to be a
discrepancy between the X-ray and optical/infrared properties of the
environment of 3C~356, which we will address later.

\section{The nature of galaxies `a' and `b'}

\begin{figure*}
\includegraphics[width=0.22\textwidth]{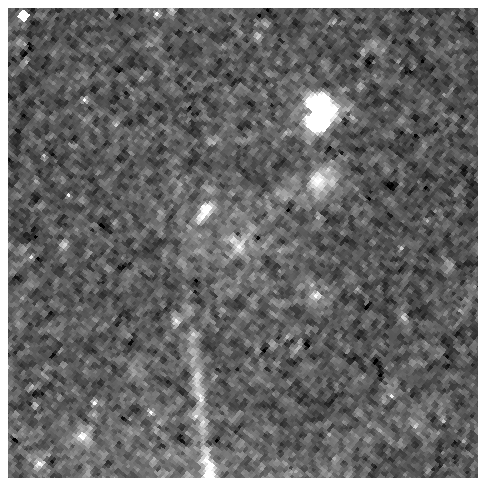}\hfill
\includegraphics[width=0.22\textwidth]{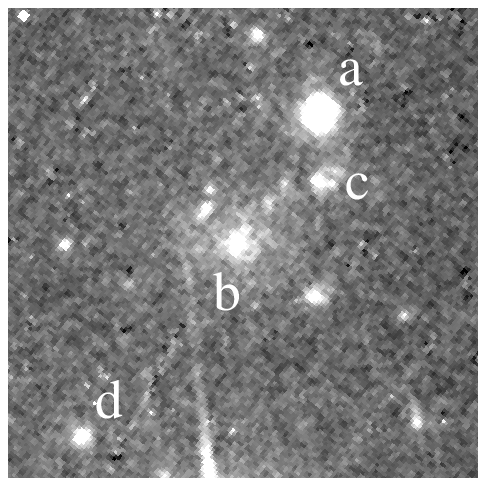}\hfill
\includegraphics[width=0.22\textwidth]{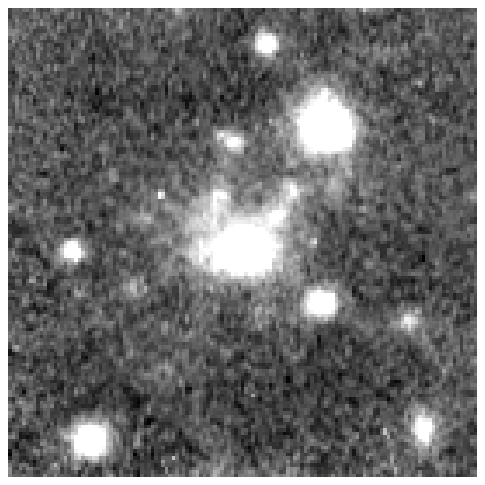}\hfill
\includegraphics[width=0.22\textwidth]{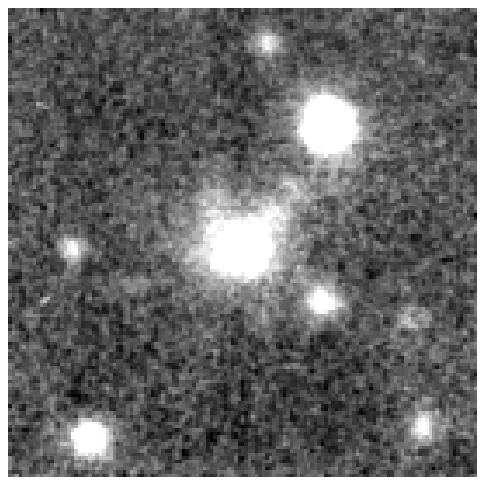}
\caption[]{Enlargements of a $15'' \times 15''$ region around 3C~356b
in (from left to right), the F622W, F814W, $J$, and $K'$ filters. The
images have all been smoothed with a 0\farcs15 (FWHM) Gaussian to enhance
low-surface brightness features, but are not at the same resolution. The
four optical components identified by Lacy \& Rawlings (1994) are indicated
in the F814W image. The linear feature extending from the bottom of the two
{\it HST\/} images is a diffraction spike from the bright star south of
3C~356.}
\label{fig:closeup}
\end{figure*}

As mentioned in Section~1 there has been considerable debate
concerning whether galaxy `a' or galaxy `b' hosts the jet-producing
AGN responsible for feeding the large-scale radio emission. We
reconsider the location of the radio source's central engine in this
section, and to aid the discussion we show close-ups of the region
around the two galaxies in all four filters in Fig.~\ref{fig:closeup}.

\begin{figure}
\includegraphics[angle=-90,width=\colwidth]{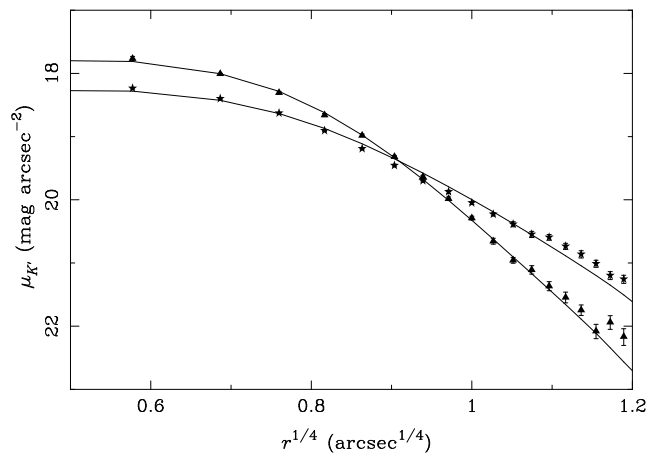}
\caption[]{Radial surface brightness profiles of components `a'
(triangles) and `b' (stars) in the $K'$-band, together with best fits
(solid lines), determined from the two-dimensional data.}
\label{fig:radprof}
\end{figure}

We first consider the morphologies of both galaxies. Radial surface
brightness profiles were constructed in the $K'$ band and fit using the
two-dimensional routine of Simpson, Ward \& Wall (2000). The results are
shown in Fig.~\ref{fig:radprof}, and we obtained half-light radii of
0\farcs36 (2.9\,kpc) and 1\farcs61 (13.1\,kpc) for `a' and `b',
respectively. In neither case was a significant nuclear excess required to
obtain a good fit. Both galaxies lie on the Kormendy relation for massive
ellipticals after accounting for passive evolution in the manner of Best,
Longair \& R\"{o}ttgering (1998), and, assuming no further merger activity,
will evolve into $\sim 3L^*$ galaxies by the present epoch. Our radius for
`a' is much smaller than the 0\farcs9 determined by Best et al.\ (1998); we
attribute their result to their much poorer seeing and the use of a
one-dimensional fitting algorithm which is extremely sensitive to errors in
the sky level and companion objects, and improperly weights the data (see
Simpson et al.\ 2000 and Taylor et al.\ 1996 for more details). The profile
of `b' has an excess at radii $\ga 1\farcs5$, similar to that of a cD
galaxy; however, a visual inspection of our image clearly reveals that this
excess is not azimuthally-symmetric and is caused by filamentary-like
structures emanating from galaxy `b'.

\subsection{The location(s) of the AGN(s)}

While Lacy \& Rawlings (1994) favoured `b' as the source of the radio jets
and suggested that `a' was the site of a jet--cloud interaction, later
spectropolarimetric observations have revealed scattered broad line
emission from `a', and energetic arguments indicate that these lines
originate locally (Cimatti et al.\ 1997). Galaxy `a' is therefore a true
AGN, whose optical polarization has its electric vector position angle
accurately perpendicular to the axis of the galaxy's `dumbbell' structure
(Best et al.\ 1997).  The dumbbell axis is in turn roughly aligned with the
radio axis and, of course, the vector joining `a' and `b'. Together with
the high optical polarization of `a', these facts suggest that much of the
blue emission from `a' is reflected from a luminous buried quasar; this
also explains the much bluer colour of object `a' ($R-K' = 4.1$, compared
to $R-K'=5.6$ for `b'). The discovery of weak optical polarization with a
broadly similar electric vector position angle in `b' could mean that some
of this light is being scattered by material associated with `b' (Cimatti
et al.\ 1997). The low excitation [O{\sc~ii}] line emission in `b' could
also plausibly be excited by the quasar in `a'.

It is the nature of galaxy `b' which is therefore unclear, and we
investigate further here. We first briefly revisit the three arguments
of Lacy \& Rawlings (1994) to favour `b' as the radio galaxy. One was
based on the $K$-band magnitudes of the two objects. According to
Eales \& Rawlings' (1990) photometry, `b' lay on the $K$--$z$ relation
for 3C radio galaxies, while `a' was too faint. More recent data
indicate that the magnitudes of components `a' and `b' in standard
63.9-kpc apertures are $16.98\pm0.05$ and $16.71\pm0.04$, respectively
(from the UKIRT data of Simpson et al.\ 1999; the increased accuracy
of our Subaru data is outweighed by the uncertainty in the $K'-K$
colours), with a small ($\sim 7$\%) amount of emission-line
contamination to the broad-band magnitude of `a'. Both are therefore
consistent with the $K$--$z$ relation for powerful radio galaxies
(e.g.\ Simpson et al.\ 1999).  Lacy \& Rawlings (1994) also noted the
1\farcs5 displacement of `a' from the line joining the radio hotspots,
but this is not unusual among powerful radio galaxies (e.g., 3C~55 and
3C~265 from Fernini et al.\ 1993). The remaining argument of Lacy \&
Rawlings, which we study in more detail here, is the flat radio
spectrum of `b' compared to the steep spectrum of `a' (Fernini et al.\
1993).

Powerful jet-producing AGN have traditionally been believed to have
flat-spectrum radio cores so the spectrum of `b' seems typical whilst that
of `a' between 5 and 8.4\,GHz seems unusually steep ($\alpha=1.1$, $S_\nu
\propto \nu^{-\alpha}$; Fernini et al.\ 1993). However, Athreya et al.\
(1997) studied 12 powerful high-redshift ($z > 2$) radio galaxies and found
that three have cores at least as steep as that of galaxy `a', Even
allowing for the slightly higher rest-frame frequencies probed by their
study, this no longer seems to be a strong argument against having `a' as
the source of the radio jets. particularly as its spectrum flattens
dramatically below 5\,GHz ($\alpha^5_{1.5}<0.3$; Fernini et al.\ 1993).

Having negated the arguments of Lacy \& Rawlings (1994) against `a' being
the core of the radio source, we note that the observed arm-length
asymmetry in 3C~356 favours it. If `a' is the radio galaxy, the fractional
separation distance (Bahati 1980) is $x=0.19$, while if `b' is the correct
ID, $x=0.34$. In the sample of 95 FR\,II 3CRR radio sources (72 radio
galaxies and 23 quasars) studied by Best et al.\ (1995), only 4 radio
galaxies and 2 quasars have $x>0.34$, while there are 14 radio galaxies and
9 quasars with $x>0.19$. We find similar numbers from the slightly larger
sample of Arshakian \& Longair (2000). We will discuss the arm-length
asymmetry further in Section~\ref{sec:timescales}.

We also note that our new near-infrared spectroscopy clearly places `a' on
the [O{\sc~iii}]--radio luminosity correlation of Jackson \& Rawlings
(1997), while our limit for `b' falls far below it and is a factor of $\ga
2$ below all other upper limits. We therefore reach the opposite conclusion
to Eales \& Rawlings (1990) and Lacy \& Rawlings (1994), namely that `a'
not `b' is more likely to be the jet-producing AGN, although only a deeper
radio observation revealing jets will provide a conclusive result.

If `a' is the core of the radio source, however, we are left with the
problem of explaining the radio emission from galaxy `b', which is several
orders of magnitude more luminous than is typical of normal galaxies (e.g.\
Fabbiano et al.\ 1987). A flat spectrum suggests optically-thin thermal
emission from a starburst, but the brightness temperature at a rest-frame
frequency, $\nu_{\rm test}$, of 10\,GHz is $T_{\rm B} > 4300$\,K, which is
too high (cf.\ Condon et al.\ 1991).  Alternatively, we could be seeing the
turnover in a `normal' starburst spectrum (i.e.\ thermal plus non-thermal
emission) where the free--free optical depth is unity at $\nu_{\rm rest}
\sim 10$\,GHz. In such a case, the observed 8.4-GHz flux is approximately
equal parts thermal and non-thermal emission (equation 7 of Condon et al.\
1991), and the implied production rate of ionizing photons is therefore
$N_{\rm UV} \sim 10^{57}$\,s$^{-1}$ (Condon \& Yin 1990). This corresponds
to a star formation rate of $\sim 10^4 M_\odot$\,yr$^{-1}$ (Kennicutt
1982), far in excess of the limits from the H$\beta$ and submillimetre
emission (Archibald et al.\ 2001). A more plausible explanation is offered
from the similarity between the ultraviolet--optical spectrum of component
`b' (Cimatti et al.\ 1997 plus our CGS4 data) and that of the LINER
NGC~1052 (Gabel et al.\ 2000). Both are core-dominated, flat-spectrum radio
sources (Wrobel 1984), and NGC~1052 is an order of magnitude less luminous
in both line and radio emission, suggesting that 3C~356b is simply a
scaled-up version of the low-redshift object.  The implication is that the
radio emission, and potentially the narrow line emission, from galaxy `b'
is caused by a second, separate AGN. Although it might seem perverse to
demand the presence of a second AGN, very similar systems are known to
exist, such as the two flat-spectrum ellipticals near the centre of the
low-redshift giant radio galaxy 3C~326 (Rawlings et al.\ 1990), or the
double-nucleus radio galaxy 3C~75 (Owen et al.\ 1985) and in the next
section we describe how they may have been synchronously triggered.

\subsection{Evidence that `a' and `b' suffered an interpenetrating 
collision}
\label{sec:collision}

Our new deep near-infrared images of 3C~356 (Fig.~\ref{fig:closeup})
confirm the impression from the {\it HST\/} data (e.g.\ Best et al.\ 1997)
that galaxy `b' is highly disturbed; Best et al.\ suggest that this
disturbance is caused by an interaction with the southern radio jet from
`a'.  We argue here that an attractive alternative explanation is that `a'
and `b' suffered an interpenetrating collision around the time the jets
from 3C~356 were first triggered.

\subsubsection{Nature of the debris}

Visual inspection of Fig.~\ref{fig:closeup} shows that the 3C356 `a' and
`b' system is very reminiscent of a pair of colliding galaxies (see, e.g.,
fig.~3 of Struck 1999).  A `bridge' can clearly be seen pointing from `b'
towards `a', and there is additional emission to the north and east of `b',
including an arc-like feature which is bright in the optical images and a
filament arcing northwards.  Component `c' (detected in [O{\sc~ii}] by Lacy
\& Rawlings 1994) is also between the two main galaxies.

\begin{figure}
\includegraphics[angle=-90,width=\colwidth]{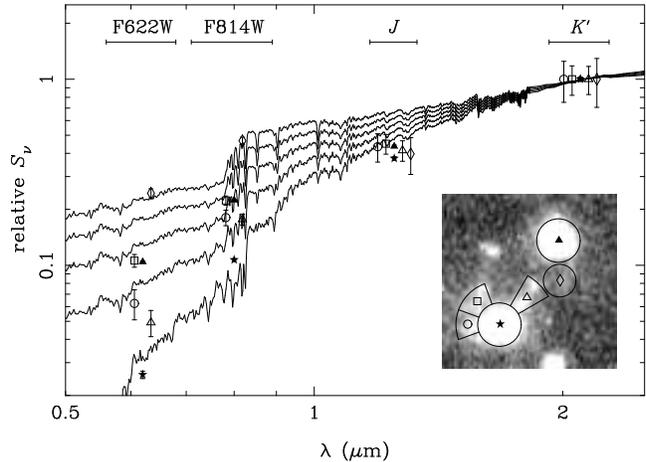}
\caption[]{Four-colour photometry of six regions in the 3C~356 system
including components `a' (filled triangle), `b' (filled star), `c'
(open diamond) and three regions of debris. The inset shows these
regions superimposed on the $J$-band image. The photometry has been
normalized to the $K'$ flux, and some of the points have been slightly
offset horizontally for clarity.  Overplotted are the spectra (at
$z=1.08$) of a 4.3-Gyr-old elliptical plus a 50-Myr-old instantaneous
burst which contributes (from bottom to top) 0,10,20,30,40\% of the
$K'$-band light.}
\label{fig:debris}
\end{figure}

In Fig.~\ref{fig:debris} we plot the four-colour broad-band SEDs of four
regions of debris, plus the galaxies `a' and `b'. These are compared to the
SEDs (from Fioc \& Rocca-Volmerange 1997) of a 4.3-Gyr-old E galaxy (a good
fit to galaxy `b') combined with varying fractions of a 50-Myr-old (see
Section~\ref{sec:timescales}) instantaneous starburst (note that the
starburst model has an essentially flat spectrum and could therefore be
equally representative of scattered quasar light). A discrepancy between
the observed $J-K'$ colours and those of the models is noted, but is hard
to explain given the agreement in Fig.~\ref{fig:colcol}. With the exception
of the very blue ($R-K' \approx 3.2$) component `c', the rest of the debris
is well-modelled by a combination where only 5--20 per cent of the
$K'$-band light (and hence 1--5 per cent of the stellar mass) originates
from young stars (if the normalization is made to the $J$-band photometry,
this mass fraction rises to $\sim 10$\% for galaxy `a' and the region to
the northeast of `b'). It is clear, therefore, that the extended light in
the system cannot be due solely to recent star formation caused, for
example, by the passage of the radio jets, but must be predominantly
composed of old stars disturbed during the collision between `a' and `b'.

While there is remarkably little discussion in the literature of what
happens when two massive ellipticals interpenetrate, one important mode by
which material can be ripped from a galaxy is mediated by tidal torques
(e.g.\ Struck 1999) which obviously affects stars as well as gas, and so
there seems to be no physical reason why the debris cannot be material
stripped from `b' (indeed, the colours of `b' and the `bridge' are very
similar; Fig.~\ref{fig:debris}). It seems far harder to find a physical
process by which the disturbed structures of `b' can be promoted by
interactions with a radio jet. Lacy et al.\ (1998) present a comprehensive
study of a probable jet--galaxy interaction involving the radio source
3C~441, from which it seems clear that there are significant effects on gas
in a galaxy struck by a jet, but little or no effect on old stars since
there is no obvious way in which the jet power can be coupled effectively
to their motions.

Similarly, the red colours of component `d' (which are very similar to
those of `b') imply that it is an old galaxy and therefore its high
velocity relative to `a' and `b' cannot be explained by an interaction
with the radio jet in the manner proposed by Lacy \& Rawlings (1994).
Neither can it be explained as a result of a strong gravitational
interaction with the two main galaxies, since the maximum velocity
which can be attained in such a scenario is $v_{\rm max} \approx
(2G(M_{\rm ab})/r)^{1/2} \approx 1000 (M_{\rm
ab}/10^{12}M_\odot)^{1/2} (r/10\,{\rm kpc})^{-1/2}$\,km\,s$^{-1}$,
where $M_{\rm ab}$ is the combined mass of components `a' and `b', and
$r$ is the impact parameter. Neither does component `d' appear to be
disturbed, although it does possess weak [O{\sc~ii}] emission. We are
therefore left with the rather unsatisfying conclusion that it is an
unrelated background galaxy.

\subsubsection{Time-scales}
\label{sec:timescales}

Adopting the physical model for FR\,II radio sources of Willott et al.\
(1999), we can use the observed properties of 3C~356 to estimate its total
jet power $Q$ and the time $t$ since the jets were first triggered. If the
jet axis makes an angle $\theta$ with respect to the line of sight, then
equation (11) of Willott et al.\ yields $Q = 6.3 \times 10^{39} (\sin
\theta)^{31/28}$\,W, where we have adopted the following `canonical' values
for the source and environmental parameters discussed by Willott et al.:
$c_1=2.3$ and $f = 20$ (see also Blundell \& Rawlings 2000); $n_{100} =
3000\,\rm m^{-3}$ and $\beta = 1.5$. Equation (9) of Willott et al.\ gives
the age of the radio source as $t = 3.4 \times 10^7 (\sin
\theta)^{-43/28}$\,yr.

So, if the merger axis of `a' and `b' is close in three dimensions to the
jet axis (e.g.\ West 1991; see also the discussion in the next section),
then the point of closest approach happened $\sim D_{\rm a-b} / (\Delta v
\tan \theta) = 3.2 \times 10^7 (\tan \theta)^{-1}$\,yr ago, where $D_{\rm
a-b} = 39$\,kpc is the projected separation between `a' and `b', and
$\Delta v = 1200$\,km\,s$^{-1}$ is the line-of-sight velocity
difference. Thus, accounting for uncertainties like small ($\sim 30^\circ$)
misalignments between the radio axis and the velocity vector of `a', it
looks highly plausible that the jets were triggered during the
interpenetrating collision between `a' and `b', a few tens of Myr in the
past. This simultaneity requires a pre-existing supermassive black hole at
the centre of `a', in line with the suggestion of Willott, Rawlings \&
Blundell (2001) that $1 \la z \la 2$ radio galaxies are undergoing a
rebirth.

Since the fractional separation difference cannot be larger than the
expansion speed of the outer radio components, $\beta \equiv v_{\rm hs}/c
\approx 0.04$ (see Blundell \& Rawlings 2000 for a recent discussion of
expansion speeds), the large arm-length asymmetries cannot be due to light
travel time effects, and the most likely cause is an environmental
asymmetry. If so, the model of Willott et al.\ (1999) implies that the mean
density to the southeast of `a' is $\sim 4$ times higher than that to the
northwest. Radio depolarization measurements (Pedelty et al.\ 1989) also
indicate a denser medium to the southeast, and visual inspection of
Fig.~\ref{fig:kimage} suggests that this may be linked to a larger
concentration of companion galaxies to the southeast, presumably associated
with the group containing `b'.

\section{Discussion}

Our study of 3C~356 has led us to a model in which a powerful jet-producing
AGN and a weaker, but possibly also jet-producing, AGN are stimulated in
the 3C~356 system as the result of an interpenetrating collision between
two massive galaxies, which occurred along an axis closely aligned with the
radio jet axis. The high relative velocity of `a' and `b' indicate that
they are not members of the same group, while the small number of companion
galaxies suggests that the system will relax to a relatively poor cluster,
no richer than Abell class 0.  This begs two questions which we address in
this section. First, how do we square these inferences with the X-ray
properties of this system, which are more reminiscent of a richer cluster
of galaxies? And second, can the trigger mechanism inferred for 3C~356 be
generalized to other objects with powerful radio jets?

\subsection{What causes the X-ray emission from 3C~356?}
\label{sec:xray}

\begin{figure}
\includegraphics[width=\colwidth]{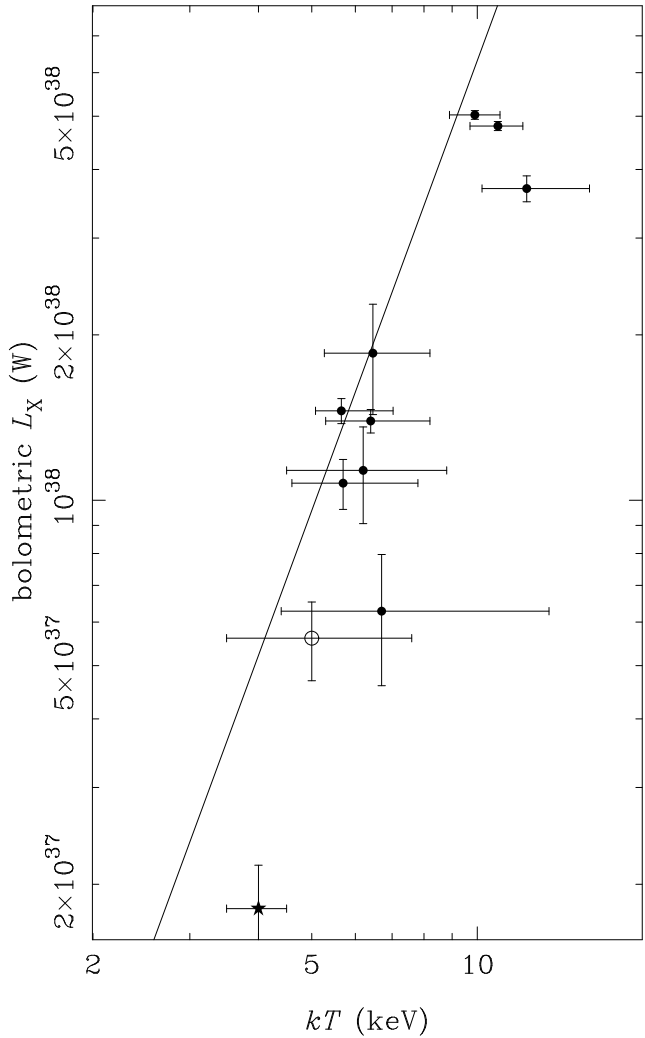}
\caption[]{Luminosity--temperature relation for distant ($z>0.5$)
clusters from Donahue et al.\ (1999), Della Ceca et al.\ (2000),
Schindler (1999), and Fabian et al.\ (2001). The star represents
3C~356 and the open circle 3C~294. The solid line is the relationship
for nearby clusters from Wu, Xue \& Fang (1999).}
\label{fig:lxtx}
\end{figure}

The \textit{ROSAT\/} images of 3C~356 reveal the X-ray emission to be
extended; while there are no significant sources in the HRI image (Crawford
\& Fabian 1996), the PSPC image (Crawford \& Fabian 1993) tentatively
suggests that the emission peaks to the north of the radio core and is
extended in the direction of the radio jets. The bolometric X-ray
luminosity of 3C~356 is $L_{\rm X} \approx 2 \times 10^{37}$\,W (Crawford
\& Fabian 1993, corrected to our cosmology), making it lie close to the
luminosity--temperature relation for nearby clusters
(Fig.~\ref{fig:lxtx}). However, as we have seen in
Section~\ref{sec:richness}, the small number of galaxies around 3C~356 does
not support the existence of a cluster with the richness implied by the
X-ray temperature.

We consider the possibility that the X-ray emission from 3C~356 is a
superposition of cooler, less luminous (i.e., consistent with the
poorness of the optical/infrared environment) intracluster gas with an
attenuated power law from the AGN in galaxy `a'. Crawford \& Fabian
(1996) estimate that no more than one-third of the observed soft X-ray
emission can arise from a point source, and this would reduce the
bolometric luminosity of the thermal component (which becomes cooler
due to the addition of the hard non-thermal emission) by a factor of
$\sim 2$, insufficient to invalidate the arguments made at the end of
Section~\ref{sec:richness}.

We therefore turn to a scenario of merging subclusters, such as that
proposed by Markevitch, Sarazin \& Vikhlinin (1999) for the cluster around
Cygnus~A. Suggestions that radio galaxy activity and cluster mergers are
related have also been advanced by Burns et al.\ (1993, 1994). Markevitch
et al.'s model consists of two hot, massive clusters in the early stages of
a merger, whereas the above arguments suggest that 3C~356 represents the
late stages of a merger between two much less massive groups. Typical
properties for such groups are $n_{\rm e} \sim 10^3$\,m$^{-3}$, $kT \sim
1$--2\,keV, and core radius $a \sim 100$\,kpc (e.g.\ Kriss, Cioffi \&
Canizares 1983; Mulchaey \& Zabludoff 1998). When two clusters, each of
mass $M$, fall from a large distance to a separation of $2a$ (which can be
considered as the start of the merger), their relative velocity is $\Delta
v = ({2GM/a})^{1/2} \approx 1000(kT/1\,{\rm keV})^{1/2}$\,km\,s$^{-1}$
(i.e.\ very similar to the current line-of-sight velocity difference
between the two main galaxies). From equations 1 and 2 of Markevitch et
al.\ (1999), this results in post-shock gas with twice the temperature and
slightly more than twice the density of the pre-shock material, and hence
almost an order of magnitude higher emissivity (these numbers are
insensitive to the exact value of the initial gas temperature). For a
merger as advanced as that of 3C~356, where the two central galaxies have
already suffered an interpenetrating collision, it is reasonable to assume
that approximately half the gas from each group has passed through the
shock front, and hence the X-ray luminosity of 3C~356 will be about an
order of magnitude more than that of the original galaxy groups. 
Since emission from thermal bremsstrahlung has a bolometric luminosity $L
\propto n_{\rm e}^2 T^{1/2}$, the merger causes 3C~356 to simply move up the
luminosity--temperature relation, $L_{\rm X} \propto T_{\rm X}^{2.72}$ (Wu
et al.\ 1999), and its X-ray properties will therefore resemble those of a
single, richer cluster (this statement is true for a wide range of merger
parameters). The post-shock gas has a cooling time-scale, $t_{\rm ff} \sim
3n_{\rm e} kT / \Lambda \sim 10^{10}$\,yr, much longer than the shock
crossing time, $t_{\rm sh} \sim a/\Delta v \sim 10^8$\,yr, and it will
therefore dominate the observed X-ray emission due to its much higher
surface brightness. Higher-quality X-ray data could either spatially or
spectrally separate the pre- and post-shock material and are obviously
desirable.

This model can also explain the alignment effect in 3C~356, as well as the
tentative alignment between the X-ray-emitting gas and the radio axis.
These alignments are all secondary and arise because the radio emission,
X-ray emission, and galaxy interaction direction are all aligned with the
direction of the cluster merger. The X-ray--merger alignment is the easiest
to understand, since the X-rays arise from the gas travelling in the
direction of the merger which has passed through the shock front. The line
joining `a' and `b' will also be aligned with this direction since the
collision velocity between the two merging groups is larger than the
velocity dispersion of either group and so the galaxies will approach each
other along a vector which is biased towards the merger axis. To explain
the alignment of the radio jets, we turn to West (1994), who suggested that
supermassive black holes form with their spin axes aligned with the local
large scale structure, as a result of a series of highly anisotropic
mergers. Since the interaction between `a' and `b' is unlikely to have
caused any precession of the spin axis of the jet-producing black hole in
`a' (Scheuer 1992), when this dormant black hole is re-fuelled (see
Section~\ref{sec:timescales}), the radio jets will align with the filament
of large scale structure, along which the groups containing `a' and `b' are
presumably merging.

We also consider the possibility that the X-ray emission is boosted in a
more direct way by the radio source activity. First, as Fabian et al.\
(2001) note in the case of the $z=1.786$ radio galaxy 3C~294,
inverse-Compton scattering of CMB photons by low-$\gamma$ electrons can
produce luminous X-rays. This mechanism requires bubbles of electrons which
are spatially separated from the ultra-relativistic material visible as
high-frequency radio emission, but studies of M~87 (Owen, Eilek \& Kassim
2000) show that this is plausible, at least in this object. In addition,
low-$\gamma$ electrons are observed to be distributed on large scales in
merging systems where it is plausible that they are re-energized by shocks
which would extend their effective lifetimes (R\"{o}ttgering et al.\
1997). As also noted by Fabian et al.\ (2001), this scenario requires two
bubbles of hot confining gas, which can be naturally associated with the
merging subclusters in our model, and therefore also predicts an alignment
between the X-ray and radio axes.

Second, Kaiser \& Alexander (1999) suggested that the X-ray emission from
3C~356 may arise from gas in the intergalactic medium which has been heated
by the passage of the expanding radio source. However, their fig.~10
indicates that the required luminosity of $\sim 10^{37}$\,W cannot be
achieved for a source with a mean arm length of $\sim 300$\,kpc unless the
density profile index $\beta < 1$ (we know of no sources where such a flat
profile has been observed) and/or the IGM density is larger than their
assumed value, which is typical of 3C radio source environments at $z \sim
1$ (this same value is justified by Willott et al.\ 1999). Since the
environment of 3C~356 i somewhat poorer than this, a higher density is
inconsistent with the observations. Also, Kaiser \& Alexander's model
requires a higher density on the north side of the radio source since the
X-ray emission peaks to the north of the core (Crawford \& Fabian 1993),
which is at odds with the longer arm length if the IGM is static. Our model
naturally explains this apparent discrepancy if the group approaching from
the south were denser (as we claim to be the case in
Section~\ref{sec:timescales}). This gas will be the most luminous in X-rays
after it has been compressed and heated by passing through the shock front
to the north side. On the other hand, the radio hotspots advance much more
rapidly than the slow ($\sim 10^3$\,km\,s$^{-1}$) IGM and therefore work
against the pre-shock material, which is denser to the south.

It seems plausible that some X-ray emission may arise from processes
directly associated with the radio source, but a full consideration of all
its observed properties seem to demand that 3C~356 is at the centre of a
merger between two galaxy groups. The time-scales involved suggest that the
radio source was triggered by the close approach, or interpenetration, of
the dominant galaxies from each group, which channelled fuel onto a dormant
supermassive black hole. Our observations and model are at odds with the
idea of Bremer et al.\ (1997) that powerful radio galaxies lie at the
centres of rich clusters with strong cooling flows. However, the richness
of the cluster around 3C~356 is fairly typical of powerful radio sources at
high redshift, and it is therefore worth investigating whether our model
for 3C~356 can be applied to other radio-loud AGN.

\subsection{Our model and 3C~294}

The radio galaxy 3C~294 has recently been observed at high resolution in
X-rays with {\it Chandra\/} (Fabian et al.\ 2001) and at near-infrared
wavelengths with Adaptive Optics on the Canada--France--Hawaii and Keck
Telescopes (Stockton, Canalizo \& Ridgway 1999; Quirrenbach et al.\
2001). The excellent quality of these data make it suitable for testing the
generality of our model.

Quirrenbach et al.'s observations reveal two galaxies separated by $\sim
8$\,kpc, the brighter of which has a diffuse, clumpy morphology, while the
fainter is compact and is identified with the source of the radio
jets. Applying the model of Willott et al.\ (1999) to 3C~294, we derive a
jet power $Q = 1.5 \times 10^{40} (\sin \theta)^{31/28}$\,W, and an age of
$t = 4.1 \times 10^6 (\sin \theta)^{-43/28}$\,yr. This much younger age is
consistent with the smaller separation of the two optical components in
3C~294, and the (rest-frame) optical properties of this system are then
completely analogous to those of 3C~356 at an earlier epoch. It is also
worth noting that in both 3C~356 and 3C~294, the radio source is identified
with the more compact object, while it is the other galaxy which is larger
and more luminous, and apparently more like a brightest cluster
galaxy. This appears to be at odds with Best et al.'s (1998) suggestion
that $z \sim 1$ radio galaxies are (the progenitors of) cD galaxies unless
the galaxy systems are always destined to merge.

The X-ray observations of 3C~294 also support a model of merging clusters.
The high resolution of \textit{Chandra\/} reveals double-lobed
(`hour-glass' shaped) emission, not typical of a relaxed rich cluster, and
more reminiscent of $N$-body simulations of merging clusters (Roettiger,
Burns \& Loken 1993). Our interpretation is that we are observing gas which
has been heated by two shocks propagating in opposite directions from the
site of the collision. As discussed above, this produces hot, bright X-ray
emission along the merger direction, surrounded by cooler, fainter,
unshocked gas from the original clusters. As with 3C~356, a collision
velocity of $\sim 1000$--2000\,km\,s$^{-1}$ could raise the temperature
from $\sim 2$\,keV to the observed 5\,keV, and would result in shocks
propagating at $\sim 500$\,km\,s$^{-1}$ relative to the collision site. If
the two components observed by Quirrenbach et al.\ (2001) were at the
centres of the two groups, each shock would have travelled a distance of
$\sim 100$\,kpc, consistent with the extent of the X-ray lobes. Since the
radio source is not triggered until much later, it is possible for the
shocked gas to have a larger extent than the more rapidly expanding radio
lobes. Again, the major axis of the X-ray emission is broadly aligned with
the radio axis, as predicted in our model (see Section~\ref{sec:xray}). The
lack of alignment between the radio axis and the line joining the two
bright galaxies can be explained by the youth of the radio source, since we
are seeing the two components soon after their time of closest approach.

We use the X-ray data of 3C~294 to draw two inferences which are relevant
in our search for a single trigger mechanism for powerful radio sources at
these intermediate epochs. First, there is no evidence for an enhancement
of the X-ray emission at the edge of the radio cocoon, as predicted by
Kaiser \& Alexander's (1999) model for heating of the IGM by the radio
source. Second, the highly-peaked X-ray surface brightness which is the
signature of a cooling flow core is not present (the weak X-ray core
coincident with the radio galaxy is consistent with a heavily-absorbed
power law), and therefore we do not believe that 3C~294 lies at the centre
of an established cooling flow. Our model of merging clusters therefore
provides the only way to explain the properties of both 3C~356 and 3C~294.

\subsection{Our model and Cygnus~A}

Our model may also be applicable to the most luminous $z<0.5$ radio source,
Cygnus~A. Cyg~A is the brightest galaxy in a cluster which is merging with
a richer cluster (Owen et al.\ 1997; Markevitch et al.\ 1999) along an axis
aligned with the radio jets. This is a qualitatively similar situation to
that we have proposed for 3C~356 and 3C~294, and West (1994) noted this
alignment as support for his model. The quantitative differences can be
easily explained. First, the much richer environment is an inevitable
consequence of hierarchical merging (Eke, Navarro \& Frenk 1998), since
more massive clusters will have had time to form by the later epoch.
Second, the earlier stage of the cluster merger at which the radio source
is created can be explained by the higher galaxy density, since it is a
galaxy--galaxy interaction which provides the trigger. While there is only
tentative evidence for a galaxy merger in Cyg~A (see Stockton, Ridgway
\& Lilly 1994), there is recent star formation (Jackson, Tadhunter \&
Sparks 1998) whose (admittedly highly uncertain) age is consistent with the
age of the radio source ($\sim 6.4$\,Myr). Although very violent
interactions are responsible for triggering the radio sources 3C~356 and
3C~294, the weaker interaction in the case of Cyg~A may still be sufficient
to trigger nuclear activity, since fuel is likely to be plentiful as a
result of the massive cooling flow ($\dot{M} \sim 250 M_\odot$\,yr$^{-1}$;
Reynolds \& Fabian 1996). Unlike at $z \sim 1$, where the interaction must
provide fuel \emph{and\/} drive it into the nucleus of the active galaxy,
Cyg~A only requires a relatively small gravitational perturbation to
disturb the reservoir of gas which is already present. We note that the
weaker, yet still extremely powerful, radio source 3C~295 also lies at the
centre of a massive cooling flow (Allen et al.\ 2001) and appears to be
undergoing an interaction (Smail et al.\ 1997) which could provide fuel to
power the nuclear accretion. We therefore offer support for the model of
Bremer et al.\ (1997) when applied to low-redshift sources where cooling
flows have been set up, and suggest that the unspecified trigger mechanism
is a galaxy--galaxy interaction.

\subsection{The trigger mechanism(s) for powerful radio sources}

A powerful radio source requires (i) a supermassive, presumably spinning,
black hole; (ii) available fuel; and (iii) a mechanism for delivering this
fuel to an accretion disc around the black hole.

The formation of the black hole is likely to be accompanied by a period of
AGN activity, as fuel is plentiful. This is believed to be the evolutionary
state of the highest-redshift ($z \ga 3$) radio sources. Since black holes
are not destroyed, lower-redshift radio sources require only fuel and a
suitable delivery mechanism to be triggered in a suitable host galaxy. At
intermediate redshifts ($1 \la z \la 2$), elliptical galaxies do not have
large internal reservoirs of gas, and our study of 3C~356 and 3C~294
suggests that fuel can be made available and delivered by a violent
galaxy--galaxy interaction which is most likely to be orchestrated by the
merger of two groups or clusters of galaxies. We suggest that this is
likely to be a very important, potentially dominant, trigger mechanism for
powerful radio sources at these cosmological epochs. At even lower
redshifts ($z \la 1$), fuel may become readily available from a cooling
flow (Bremer et al.\ 1997), and therefore a less violent trigger is
required to perturb the gas onto the black hole.

The steep decline in the number density of extragalactic radio sources from
$z \sim 1$ to the present epoch (e.g.\ Dunlop \& Peacock 1990; Willott et
al.\ 2001) can be understood in terms of the currently-favoured low
matter-density Universe as the end of the era of cluster formation via
group--group mergers, and hence the end of the epoch of violent
interactions which can fuel dormant black holes (see also Blundell \&
Rawlings 1999). It is worth pointing out that this is precisely the
opposite conclusion to that reached by Fabian \& Crawford (1990), who
suggested that cluster mergers \emph{turned off\/} quasars since it was
then believed that $\Omega_{\rm m}=1$ and hence massive clusters formed at
a much later epoch.

Our model, if applicable to intermediate-redshift radio sources in general,
has an important implication for the use of radio sources to locate
clusters at $z \ga 1$, as it predicts that these radio-selected clusters
are likely to be merging systems. These systems are atypical and cannot be
used to infer details about the evolution of the cluster X-ray luminosity
function. However, proving that these systems are not simply relaxed
clusters is likely to be difficult since the X-ray luminosity and
temperature of these systems mimic those of a single, richer cluster (see
Section~\ref{sec:xray}). In addition, while the merger directions of radio
galaxies lie close to the plane of the sky and therefore should produce
elongated X-ray emission, radio-loud quasars lie in systems which are
merging along axes close to the line of sight, and projection effects will
make it much harder to observe evidence of a merger. Even some radio
galaxies may be produced by mergers close to the line of sight, due to a
combination of misalignments between the axes cluster and galaxy mergers,
and between the axes of the galaxy merger and the radio emission.

\section{Summary}

We have demonstrated with new deep near-infrared imaging the absence of a
rich cluster around the $z=1.08$ radio galaxy 3C~356, apparently at odds
with the observed X-ray emission. We have proposed that the X-rays arise
from shocked gas at the interface of two merging galaxy groups and that the
two components of 3C~356 suffered a strong interaction during this merger,
stripping material from one or both galaxies and funnelling it into the
nucleus of `a', which we believe is the source of the radio jets (although
we believe that `b' is also an AGN). We have proposed an identical scenario
for the $z=1.786$ radio galaxy 3C~294, and have suggested that this
mechanism may be the dominant trigger for radio sources at $1 \la z \la
2$. We have suggested that less violent galaxy interactions may trigger
lower-redshift radio sources where a cooling flow has set up a reservoir of
fuel in the host galaxy. This model explains the rapid decline in the
number density of these objects since $z \sim 1$, and also implies that
galaxy clusters located by using radio sources will be an extremely biased
sample inappropriate for learning about cluster formation and evolution.
It is vital to establish the ways in which the radio source can directly
influence the observed X-ray properties, and also to understand how the
triggering of a radio source may be intimately related to the evolution of
its environment.

\section*{Acknowledgments}

We wish to thank Roberto de Propris for giving up valuable observing time
to obtain the CGS4 spectra, and Adam Stanford and Curtis Struck for useful
discussions. We are also grateful to Katherine Blundell for providing the
radio map in Fig.~1, and the referee, Paul Alexander, for his comments.
This work is based in part on data collected at Subaru Telescope, which is
operated by the National Astronomical Observatory of Japan, and in part on
observations with the NASA/ESA {\it Hubble Space Telescope\/}, obtained
from the data archive at the Space Telescope Science Institute, which is
operated by AURA, Inc.\ under NASA contract NAS5--26555. The United Kingdom
Infrared Telescope is operated by the Joint Astronomy Centre on behalf of
the U. K. Particle Physics and Astronomy Research Council.

\end{document}